\documentclass[superscriptaddress,nofootinbib,preprintnumbers,showpacs,tightenlines,notitlepage]{revtex4}
\usepackage[latin1]{inputenc}
\usepackage{graphicx}
\usepackage{amssymb}
\usepackage{color}
\usepackage{float}
\usepackage{amsmath}
\usepackage{amsfonts}
\usepackage{dcolumn}
\usepackage{hyperref}
\usepackage{amsthm}
\usepackage{color}

\def\3nab{\tilde{\nabla}}

\def\be {\begin{equation}}
\def\ee {\end{equation}}
\def\ba {\begin{eqnarray}}
\def\ea {\end{eqnarray}}
\newtheorem{thm}{Theorem}

\newcommand{\barray}{\begin{array}}
\newcommand{\earray}{\end{array}}

\newcommand{\mb}{\overline{m}}
\newcommand{\ov}{\overline}

\begin{document}

\title{ 
4-dimensional spacetimes from 2-dimensional conformal null data}
\author{Rituparno Goswami}
 \email{goswami@ukzn.ac.za}
  \affiliation{Astrophysics and Cosmology Research Unit, School of Mathematics, Statistics and Computer Science, University of KwaZulu-Natal, Private Bag X54001, Durban 4000, South Africa.}
 \author{George F. R. Ellis}
 \email{george.ellis@uct.ac.za}
 \affiliation{Department of Mathematics and Applied Mathematics and ACGC, University of Cape Town,
Cape Town, Western Cape, South Africa.}

 \begin{abstract}
In this paper we investigate whether the holographic principle proposed in string theory has a classical counterpart in  general relativity theory. We show that there is a partial correspondence:  at least in the case of vacuum Petrov type D spacetimes that admit a non-trivial Killing tensor, which encompass all the astrophysical black hole spacetimes, there exists a one to one correspondence between gravity in bulk and a two dimensional classical  conformal scalar field on a null boundary. 
\end{abstract}


\maketitle
\section{Introduction}
There has been a lot of excitement in recent times about the holographic principle in gravitational theory  \cite{Sus95}, which broadly speaking 
is the idea that boundary data uniquely determines bulk properties:
\begin{quote}
	\textbf{Null Boundary Principle 1}: \textit{The description of a volume of space can be thought of as encoded on a boundary of the region, and particularly on a light-like boundary}.
\end{quote}
This is usually developed in the context of quantum field theory and the AdS/CFT correspondence \cite{Bou02}, where a gauge theory (a conformal quantum field theory, for example super-Yang Mills theory) on its 4 dimensional boundary determines spacetime geometry in a 5-dimensional  Anti-de Sitter bulk: 
\begin{quote}
	\textbf{Null Boundary Principle 2}: \textit{The data needed to specify a spacetime geometry in the bulk satisfies a  
	conformal field theory (CFT) on the null boundary.}
\end{quote} This then leads to limits on the data that is needed to construct the spacetime when one takes the 
state of the bulk on each slice to be fully described by data not exceeding A bits, where A is the area of
the boundary of the slice \cite{Bou02}. A very fruitful interaction of this program with some other branches of physics has developed  \cite{Nat14,Hub15}.\\

Now 
in the case of general relativity, 
Null Boundary Principle 1 sounds just like 
 the null initial value problem developed by Sachs, Penrose, \textit{et al} in the 1960s (\cite{Sac62,Pen63,Pen80}).  This theory shows how data on a null 3-surface $\cal{S}$ is initial data for the standard Einstein field equations, determining the geometry of spacetime in the 4-dimensional bulk $\cal{V}$.  That data has to satisfy a set of initial value equations $\cal{I}(\cal{S})$ that hold on $\cal{S}$ (\cite{Ray}). The question that naturally arises is whether these equations $\cal{I}$ for the data on $\cal{S}$ might in fact be a form of classical conformal field theory. Then the classical counterpart of Null Boundary Principle 2 would be satisfied in a straightforward way in general relativity theory. It would of course not have the many other properties entailed by the AdS/CFT equivalence in string theory \cite{Nat14,Hub15}.\\ 
 
 In this paper, we consider classical General Relativity Theory (GRT) for a vacuum, and consider if there might be a  boundary surface/bulk (volume) correspondence exhibiting  the following feature:
\begin{quote}
	\textbf{Null Boundary Principle 3:} 
	\textit{There exists a null surface $\cal{N}$ in  spacetime and a classical conformal field theory (CCFT) without redundant degree of freedom, which generates the volume geometry $\cal{V}$ from the CCFT initial data on the boundary null surface of a vacuum spacetime.}
	\end{quote}

We show here that this is indeed the case for vacuum Petrov type D spacetimes that admit a non-trivial Killing tensor (for example the exterior spacetime for a rotating Black hole \cite{Kerr} or LRS-II vacuum spacetimes  \cite{Ell676,ElsEll96}), by  providing the initial data for the wave equation for the Lanczos potentials \cite{Lan38} on null hypersurfaces in these spacetimes. Due to the presence of an extra symmetry in terms of a non-trivial Killing tensor $K_{ab}$ that solves the Killing equation $\nabla_{(a}K_{bc)}=0$, there exist an extra symmetry in the wave equation in these spacetimes that allows separation of variables in four dimensional spacetimes \cite{frolov}. Using this we show that on a null hypersurface the initial data indeed represents a lower dimensional classical conformal field theory. \\

This {\it classical holography} differs from the quantum counterpart firstly in fact that the two dimensional conformal field on the null boundary is a classical scalar field and has Euclidean signature, so for example the idea of a central charge does not arise,  and secondly because the spacetime satisfies the vacuum Einstein equations. By the null initial value theorem for general relativity, one can in principle construct an one to one correspondence between this conformal scalar field in the boundary to the properties 
of gravity in the bulk. This is done in the following way: The Lanczos potentials in general relativity can be treated on the same footing as the electromagnetic potentials. By specifying the Lanczos potentials on a null boundary as an initial data, which we show in this paper to be a lower dimensional conformal classical scalar field, we can determine the potentials in the bulk. Then these potentials can be used to construct the Weyl tensor 
 that contains all the dynamic 
 informations of gravity for a vacuum spacetime. \\

Another important difference between the usual quantum holography and this classical counterpart is that the necessity of a negative cosmological constant disappears in the classical case considered here. Any vacuum Petrov type D spacetimes that admit a non-trivial Killing tensor will suffice for the existence of this classical correspondence; it will  hold for any value of the cosmological constant (positive, negative, or zero).

\section{Weyl tensor and Lanczos potentials}

The Riemann curvature tensor can be decomposed into its irreducible subtensors in the following way \cite{Dolan}
\be\label{Riemann}
R_{abcd}=G_{abcd}+E_{abcd}+C_{abcd}
\ee
where we have 
\be\label{G}
G_{abcd}=\frac{1}{12}R(g_{ac}g_{bd}-g_{ad}g_{bc})\;,
\ee
\ba\label{E}
E_{abcd}&=&\frac12(g_{ac}S_{bd}+g_{bd}S_{ac}-g_{ad}S_{bc}-g_{bc}S_{ad})\;,\nonumber\\
S_{ab}&=&R_{ab}-\frac14g_{ab}R\;.
\ea
The tensor $C_{abcd}$ is the Weyl tensor which satisfies the same algebraic symmetries as the  Riemann tensor, namely
\be\label{C1}
C_{abcd}=C_{[ab][cd]}=C_{[cd][ab]}\;,\; C_{[abcd]}=0\;.
\ee
Furthermore this tensor is traceless, as any contraction of this tensor vanishes:
\be\label{C2}
C^a_{bad}=0
\ee
The properties (\ref{C1}) and (\ref{C2}) imply that the left dual of this tensor is equal to its right dual:
\be\label{C3}
{}^*C_{abcd}\equiv\frac12\eta_{absm}C^{sm}_{~~cd}=C^*_{abcd}\equiv\frac12\eta_{cdsm}C_{ab}^{~~sm}\;,
\ee
where $\eta_{abcd}$ is the 4-dimensional volume element.\\

Lanczos \cite{Lan38} constructed a tensor $L_{abc}$ of type (0,3), known as the Lanczos tensor, such that the Weyl tensor can be uniquely determined by the covariant differentiation of this tensor field. It has been shown
rigorously \cite{Rob89,Dolan,Ilge, Edgar} that this tensor always exists in 4-dimensions.  The algebraic properties of this tensor can be summarised in the following way. The tensor is antisymmetric in the first two indices, and furthermore we impose an algebraic gauge choice:
\ba\label{L1}
L_{abc}&=&L_{bac}\;,\\
L_{a~~c}^{~~c}&=&0\;.
\ea
These conditions are very similar to the antisymmetry conditions of the electromagnetic field tensor. However, even with these algebraic constraints, the tensor $L_{abc}$ has 16 independent components, while the Weyl tensor has only 10. Hence one has to impose 6 differential gauge conditions in the following way:
\be\label{L2}
\nabla_cL_{ab}^{~~c}=0
\ee
With the conditions (\ref{L1}-\ref{L2}), the tensor $L_{abc}$ has exactly 10 independent components and one can construct a tensor of type (0,4), with the exact symmetry properties of Weyl tensor and equate that to the Weyl tensor. Hence we have the result \cite{Dolan}
\be\label{WL}
C_{abcd}=\frac12\left[\nabla_dL_{[ab]c}-\nabla_cL_{[ab]d}-\nabla_bL_{[cd]a}+\nabla_aL_{[cd]b}
+{}^*\nabla_d L^*_{[ab]c}-{}^*\nabla_c L^*_{[ab]d}-{}^*\nabla_b L^*_{[cd]a}+{}^*\nabla_a L^*_{[cd]b}\right], 
\ee
where the double dual is defined in the usual way
\be
{}^*\nabla_a L^*_{[bc]d}=\frac14\eta_{absm}\eta_{cdpq}\nabla^{s}L^{[mp]q}\;.
\ee
The above set of 10 independent equations are known as the {\it Weyl-Lanczos} equations. We note that just like the electromagnetic potentials, the Lanczos tensor also has a gauge invariance. Any algebraic gauge transformation of the form
\be\label{gauge}
L'_{abc}\equiv L_{abc}+\xi_ag_{bc}-\xi_bg_{ac}\;,
\ee
where $\xi^a$ is an arbitrary four vector, will leave the Weyl-Lanczos equations unchanged.\\

We can now introduce the Einstein and Schouten tensors in the usual way:
\be\label{ein}
G_{ab}=R_{ab}-\frac12g_{ab}R\;,
\ee
\be\label{schouten}
J_{[ab]c}=\nabla_{[b}R_{a]c}-\frac16g_{c[a}\nabla_{b]}R\;.
\ee
Using the above two tensors the Bianchi identities
\be\label{bian}
\nabla_eR_{abcd}+\nabla_dR_{abec}+\nabla_cR_{abde}=0\;,
\ee
can be reduced to the following divergence equations:
\be\label{eindiv}
\nabla_aG^{a}_{~b}=0\;,
\ee
and
\be\label{weyldiv}
\nabla_dC_{abc}^{~~~d}=J_{abc}\;.
\ee
Using equation (\ref{WL}) in (\ref{weyldiv}), we get using the Ricci identities and considerable simplifications, the following wave equation for the Lanczos tensor
\be\label{Lwave1}
\Box L_{abc}+2R_c^{~d}L_{abd}-R_a^{~d}L_{bcd}-R_b^{~d}L_{cad}
-g_{ac}R^{ed}L_{ebd}+g_{bc}R^{ed}L_{ead}-\frac12RL_{abc}=J_{abc}. 
\ee
For vacuum spacetimes, when the Ricci tensor and Ricci scalar vanish identically, we can easily see that each of the 10 independent components of the Lanczos tensor satisfy a massless scalar field equation 
\be\label{Lwave2}
\Box L_{abc}=0\,.
\ee
Hence the components of the Lanczos tensor can be treated on the same footing as the potentials in the electromagnetic theory in vacuum, and henceforth we will denote these components as the Lanczos potentials.

\section{Lanczos potentials in Newman Penrose formalism}
We will use the Newman-Penrose formalism \cite{NewPen62,Kraetal80} to investigate the field equations in Petroc Type D geometries. We assume that the spacetime is spanned by the Newman Penrose (NP) null tetrad $(l^a,n^a,m^a,\mb^a)$ . 
The metric in terms of these tetrads is given by
\be
g_{ab}=-2l_{(a}n_{b)}+2m_{(a}\mb_{b)}
\ee
We can define the directional derivative along each of the tetrad vector fields in the following way:
\be
D\equiv l^a\nabla_a,\; \Delta\equiv n^a\nabla_a,\; \delta\equiv m^a\nabla_a,\; \overline{\delta}\equiv \overline{m}^a\nabla_a
\ee
Also the following are the NP spin coefficients [\cite{NewPen62,Kraetal80}]:
\be
\left\{\kappa,\sigma,\nu,\lambda,\tau,\rho,\pi,\mu,\epsilon,\gamma,\alpha,\beta\right\}\;.
\ee
We note here that in general the directional derivatives do not commute. The commutation relations can be written as 
\ba\label{comm1}
\Delta D- D\Delta&=&(\gamma+\ov{\gamma})D+(\epsilon+\ov{\epsilon})\Delta-(\ov{\tau}+\pi)\delta-(\tau+\ov{\pi})\ov{\delta}\, ,\\
\label{comm2}\delta D- D\delta&=&(\ov{\alpha}+\beta-\ov{\pi})D+\kappa \Delta -(\rho+\epsilon-\ov{\epsilon})\delta-\sigma\ov{\delta}\, ,\\
\label{comm3}\delta\Delta-\Delta\delta&=&-\ov{\nu}D+(\tau+\ov{\alpha}-\beta)\Delta+(\mu-\gamma+\ov{\gamma})\delta+\ov{\lambda}\ov{\delta}\, ,\\
\label{comm4}\ov{\delta}\delta-\delta\ov{\delta}&=&(\ov{\mu}-\mu)D+(\ov{\rho}-\rho)\Delta+(\alpha-\ov{\beta})\delta-(\ov{\alpha}-\beta)\ov{\delta}\,.
\ea
In the NP formalism the 10 independent components of the Weyl tensor are reduced to 5 complex scalars in the following way:
\be
\Psi_0=C_{abcd}l^am^bl^cm^d\equiv C_{1313}\;,
\ee
\be
\Psi_1=C_{abcd}l^an^bl^cm^d\equiv C_{1213}\;,
\ee
\be
\Psi_2=C_{abcd}l^am^b\mb^cn^d\equiv C_{1342}\;,
\ee
\be \Psi_3=C_{abcd}l^an^b\mb^cn^d\equiv C_{1242}\;,
\ee
\be
\Psi_4=C_{abcd}n^a\mb^bn^c\mb^d\equiv C_{2424}\;.
\ee
Similarly we can decompose the Lanczos tensor into 8 complex scalars defined as follows \cite{Dolan, Od, OdBook}: \be
L_0=L_{abc}l^am^bl^c\;\;;\;\; L_1=L_{abc}l^am^b\mb^c\;,
\ee
\be
L_2=L_{abc}\mb^an^bl^c\;\;;\;\; L_3=L_{abc}\mb^an^b\mb^c\;,
\ee
\be
L_4=L_{abc}l^am^bm^c\;\;;\;\; L_5=L_{abc}l^am^bn^c\;,
\ee
\be
L_6=L_{abc}\mb^an^bm^c\;\;;\;\; L_7=L_{abc}\mb^an^bn^c\;.
\ee
We note that these 8 complex scalars are not linearly independent, as we will see from the NP version of the {\it Weyl-Lanczos} equations \cite{EdHo97} given below :
\be\label{WLNP1}
-\frac12\Psi_0=DL_4-\delta L_0+(\ov{\epsilon}-3\epsilon-\ov{\rho})L_4+(\ov{\alpha}+3\beta-\ov{\pi})L_0+3\kappa L_5-3\sigma L_1
\ee
\be\label{WLNP2}
-\frac12\Psi_1=DL_5-\delta L_1+(\ov{\epsilon}-\epsilon-\ov{\rho})L_5-\pi L_4+\mu L_0 +(\ov{\alpha}+\beta-\ov{\pi})L_1+2\kappa L_6-2\sigma L_2 
\ee
\be\label{WLNP3}
-\frac12\Psi_1=\ov\delta L_4-\Delta L_0+(\ov\beta-3\alpha-\ov\tau)L_4 +(-\ov{\mu}+3\gamma+\ov{\gamma})L_0+3\rho L_5-3\tau L_1  
\ee
\be\label{WLNP4}
-\frac12\Psi_2=DL_6-\delta L_2+(\ov{\epsilon}+\epsilon-\ov{\rho})L_6-2\pi L_5 +(\ov{\alpha}-\beta-\ov{\pi})L_2+\kappa L_3-\sigma L_3 +2\mu L_1
\ee
\be\label{WLNP5}
-\frac12\Psi_2= \ov\delta L_5-\Delta L_1+(\ov\beta-\alpha-\ov\tau)L_5-\lambda L_4+\nu L_0 +(-\ov{\mu}+\gamma+\ov{\gamma})L_1+2\rho L_6-2\tau L_2 
\ee
\ba\label{WLNP6}
-\frac12\Psi_3=DL_7-\delta L_3+(\ov{\epsilon}+3\epsilon-\ov{\rho})L_7 +(\ov{\alpha}-3\beta-\ov{\pi})L_3-3\pi L_6 +2\mu L_2
\ea
\ba\label{WLNP7}
-\frac12\Psi_3= \ov\delta L_6-\Delta L_2+(\ov\beta+\alpha-\ov\tau)L_6-2\lambda L_5+2\nu L_1 +(-\ov{\mu}-\gamma+\ov{\gamma})L_2+\rho L_7-\tau L_3 
\ea
\ba\label{WLNP8}
-\frac12\Psi_4= \ov\delta L_7-\Delta L_3+(\ov\beta+3\alpha-\ov\tau)L_7 +(-\ov{\mu}-3\gamma+\ov{\gamma})L_3-3\lambda L_6+3\nu L_2
\ea
We can easily see that the differential constraints on the Lanczos potentials $\{L_N\}$ are included in the eight complex  Weyl-Lanczos equations.
In terms of the NP directional derivatives, the D'Alembertian is defined as \cite{frolov}
\be
\Box\equiv 2\nabla_a(-l^{(a}n^{b)}+m^{(a}\mb^{b)})\nabla_b
\ee
Therefore the scalar wave equations for all the Lanczos potentials $\{L_N\}$ for a vacuum spacetime can be written as
\be\label{Wave2}
\left[-(\Delta+\mu+\ov\mu-\gamma-\ov\gamma)D-(D-\rho-\ov\rho)\Delta +(\delta+2\beta-\tau)\ov\delta+(\ov\delta+2\ov\beta-\ov\tau)\delta\right]\{L_N\}=0
\ee
for $N = 0-7$.

\section{Initial data on null surfaces in vacuum Petrov type D spacetimes}

In this section, we develop the initial data required for a unique solution to the wave equation (\ref{Wave2}) for the Lanczos potentials in a vacuum Petrov type D spacetime that posseses a non-trivial Killing tensor. Once the potentials are known, then we can use equations (\ref{WLNP1}-\ref{WLNP8}) to derive the 5 complex Weyl scalars. Since for the vacuum spacetimes, the Weyl tensor completely determines the dynamics of gravity, hence the solution of  (\ref{Wave2}) for all the linearly independent Lanczos potentials will uniquely determine the 4 dimensional geometry of the spacetime. In  particular this will be true on any null surface.

\subsection{Intrinsic metric on a null surface}

Since the hypersurface orthogonal to real null vectors $l^a$ (or $n^a$)  contains $l^a$ (or $n^a$) as 
\begin{equation}
n^an_a=l^al_a=0, l^an_a=-1,
\end{equation} 
therefore the projection tensor onto a locally orthogonal space now has to be defined differently.
Let us now define the projection tensor $\tilde{h}_{ab}$, which projects tensors and vectors into the 2-D 
screen space orthogonal to $l^a$ (or $n^a$) as
\be
\tilde{h}_{ab}\equiv g_{ab}+2l_{(a}n_{b)}, \,\,\tilde{h}^a_a=2,\,\,\tilde{h}_{ac}\tilde{h}^c_b=\tilde{h}_{ab}, \label{ht}
\ee
\be
\tilde{h}_{ab}l^b=\tilde{h}_{ab}n^b=0.
\ee
Therefore we can write
\be
\tilde{h}_{ab}=2m_{(a}\mb_{b)}
\ee
This is the intrinsic metric on a null hypersurface and this tensor projects all the vectors and tensors onto the 2D plane spanned by $(\boldsymbol{m}, \boldsymbol{\overline{m}})$.

\subsection{Two intersecting null surfaces}
Consider two null 3-surfaces $N_1$, $N_2$ spanned by $\{l^a,m^b,\mb^c\}$ and $\{n^a,m^b,\mb^c \}$ respectively, intersecting in a spacelike 2-surface $S_2$ spanned by $\{m^b,\mb^c \}$ (see Figure 1).
\begin{figure}[h]
\centering
\includegraphics[width=0.9\linewidth]{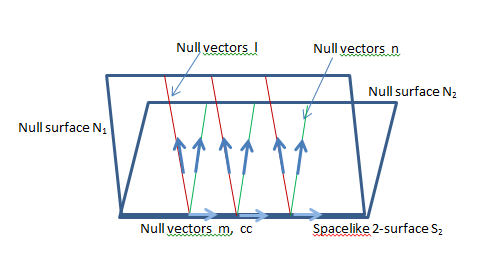}
\caption{Two null 3-surfaces $N_1$, $N_2$ intersecting in a spacelike 2-surface $S_2$.}
\end{figure}
Then data for the Einstein Field Equations on $N_1$, $N_2$, satisfying the initial value equations on these surfaces, will determine the spacetime in the 4-dimensional region lying to the future of these null surfaces. More precisely, spacetime is determined by this data in the domain $\cal{D}$ such that the causal past $J^-(p)$ \cite{HawEll73}  of every point $p$ in $\cal{D}$ intersects $N_1 \cup N_2$ and no past curve from $\cal{D}$ does not  intersect $N_1 \cup N_2$.\\

\begin{figure}[h]
	\centering
	\includegraphics[width=0.9\linewidth]{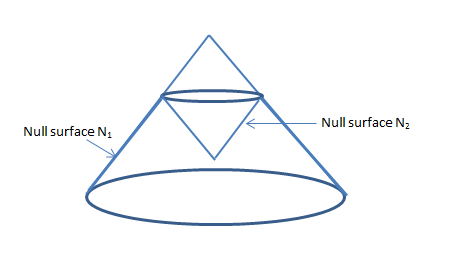}
	\caption{A null cone $N_1$ with an inverted small null cone $N_2$ to give a configuration similar to Figure 1, of two null surfaces intersecting in a spacelike 2-surface $S_2$.}
\end{figure}
This is the standard geometry for considering the null initial value problem \cite{Pen80},  as it fits well with the formalism just discussed. However we would also like to be able to set the null data on a past or future null cone, so as to connect to the way the holographic principle is sometimes stated in relation to observations. The null cone however represents a problem, because the tetrads become singular at the origin. \\

The way to handle this is shown in figure 2: add to the null cone $N_1$ with vertex point $p$, here chosen as past directed, a small future directed null cone $N_2$. Then data on $N_1 \cup N_2$ determines the geometry of the 4-dimensional spacetime 
on the full past $J^-(p)$ of the vertex point $p$ of null cone $N_1$ by determining the 4-dimensional space time from the initial data on  $N_1 \cup N_2$ (integrating to the past in the past of $N_2$ and to the future in the future of $N_2$).
 
\subsection{Vacuum Petrov Type D Spacetimes}

As discussed earlier,  
we consider vacuum  Petrov Type D spacetimes that admit a non- trivial Killing tensor $K_{ab}$ which solves the equation 
\be
\nabla_{(a}K_{bc)}=0\,. 
\ee
The Kerr spacetime, and hence as a special case the Schwarzschild spacetime,  falls in this category. This is important as this includes all the astrophysical black hole spacetimes. Both Kerr and Schwarzschild spacetimes are axisymmetric which imply that their exist a spacelike Killing vector in the $(\boldsymbol{m}, \boldsymbol{\overline{m}})$ subspace. Furthermore both metrics, being stationary,  admit a timelike Killing vector in the $ (\boldsymbol{l}, \boldsymbol{n})$ subspace too. In the case of the zero rotation, this directly relates to the Birkhoff's theorem for vacuum LRS-II spacetimes, namely that these spacetimes have an extra timelike Killing vector in the $ (\boldsymbol{l}, \boldsymbol{n})$ subspace \cite{EllGos13}.  Then we can immediately see that the non-zero spin coefficients, Weyl scalars and Lanczos potentials are not a function of the curve parameters along the integral curves of these Killing vectors. Also being Petrov type D, by the corollary of Goldberg-Sachs theorem \cite{GolSac62}, there exist  a pair of principal null congruences in these spacetimes that are geodesic and shear free. Hence in these spacetimes the following holds
\be\label{D1}
\Psi_0=\Psi_1=\Psi_3=\Psi_4=0\Leftrightarrow \kappa=\sigma=\nu=\lambda=0\;.
\ee
In these class of spacetimes it is natural to choose one of the real null vector of NP tetrad ($\boldsymbol{l}$ say) , as the tangent to one of the principal congruence. As these are geodesics, we can always make them affinely parametrised:
\be\label{D2}
Dl^a=0\Rightarrow (\epsilon+\ov\epsilon)=0\;.
\ee
Furthermore, if we make the chosen tetrad to be parallely propagated along these congruences, then we have
\be\label{D3}
\epsilon=\pi=0
\ee
Choice of such tetrad is very convenient for many calculations. Now once we have fixed the vector $\boldsymbol{l}$, there is an unique and canonical  way to choose the other real null direction $\boldsymbol{n}$ \cite{sommers}. The second null direction is defined geometrically as the unique other null
direction in which the covariant derivative of the shear-free geodetic null vector field is proportional to the shear-free geodetic field itself. This implies
\be\label{D3}
\tau=0\Rightarrow \Delta l^a= (\gamma+\ov\gamma)l^a
\ee
Choice of these two vectors ($\boldsymbol{l}$,$\boldsymbol{n}$) makes the Lie derivative of one with respect to the other, to be parallel to the shear-free geodetic null vector:
\be\label{D4}
\left[{\mathcal{L}}_{\boldsymbol{n}}\boldsymbol{l}\right]^a= (\gamma+\ov\gamma)l^a
\ee
We can always choose the coordinate in $\boldsymbol{l}$ direction, as the affine parameter of the null congruence, ($v$ say). Then the solution of the Ricci identity
\be\label{D5}
D\rho=\rho^2
\ee
is given by
\be
\rho=-\frac{1}{v+i\rho_0}\;,
\ee
where $\rho_0$ is the constant of the integration along the congruence and the origin of the coordinate $v$ is fixed by the requirement that $\rho_0$ is real. Without loss of generality we can always take it to be zero. This implies the spin coefficient $\rho$ is a real function and hence
\be\label{D6}
\rho-\ov\rho=0\;.
\ee
Using equations (\ref{D1}-\ref{D6}), we get the following 
\be\label{D7}
\nabla_{[b}l_{a]}=-(\alpha +\ov\beta)l_{[a}m_{b]} -(\ov\alpha +\beta)l_{[a}\mb_{b]} 
\ee
By tetrad rotation we can always make $(\alpha +\ov\beta)$ vanish \cite{Kraetal80}, and therefore with this choice of tetrad we have $\nabla_{[b}l_{a]}=0$, which imply the vector $\boldsymbol{l}$ is equal to a gradient of an complex function $\zeta$:
\be \label{D8}
l_a=\nabla_a \zeta
\ee
Hence we can easily see that both the null vectors $\boldsymbol{l}$ and $\boldsymbol{n}$ are orthogonal to the 2-surface described by the co-ordinates $(\zeta,\ov\zeta)$.
Because of  the relation (\ref{D4}), the other real coordinate $u$ (along the integral curves of $\boldsymbol{n}$) can always be constrained as follows \cite{sommers}
\be\label{D8}
\Delta u=1\;.
\ee
However, there remains a freedom of adding any real function of ($\zeta,\ov\zeta$) to this definition of $u$. Therefore the coordinate system thus defined, $(v,u,\zeta,\ov\zeta)$, covers the stationary vacuum spacetime and the null vectors $\boldsymbol{l}$ and $\boldsymbol{n}$ are tangent to the surfaces of constant ($\zeta,\ov\zeta$)).\\

Due to these symmetry properties of Petrov type D spacetimes, we also have a reduced set of Lanczos potentials in this case \cite{Od, OdBook, Dolan}:
\be
L_0=L_4, L_1=L_5, L_2=L_6, L_3=L_7\;.
\ee
 
With the symmetries of vacuum  Petrov Type D spacetime, the {\it Weyl-Lanczos} equations (\ref{WLNP1}-\ref{WLNP8}) becomes highly simplified and we can write them as two sets of equations. The first set evolves along $l^a$ and can be written as,
\ba
(D-\delta-\rho+2\beta)L_0&=&0, \label{WLNPa}\\
(D-\delta-\rho)L_1+\mu L_0&=&0,\label{WLNPb}\\
(D-\delta-\rho-2\beta)L_2+2\mu L_1&=&-\frac12\Psi_2, \label{WLNPc}\\
(D-\delta-\rho-4\beta)L_3+2\mu L_2&=&0. \label{WLNPd}
\ea
While the second set evolves along $n^a$ and are given by
\ba
\left[ -\Delta+\ov{\delta}-4\alpha+(-\ov\mu+3\gamma+\ov\gamma)\right]L_0+3\rho L_1&=&0,\label{WLNPe}\\
\left[-\Delta+\ov{\delta}-2\alpha+(-\ov\mu+\gamma+\ov\gamma)\right]L_1+2\rho L_2&=&-\frac12\Psi_2,\label{WLNPf}\\
\left[-\Delta+\ov{\delta}+(-\ov\mu-\gamma+\ov\gamma)\right]L_2+\rho L_3&=&0,\label{WLNPg}\\
\left[-\Delta+\ov{\delta}-4\alpha+(-\ov\mu-3\gamma+\ov\gamma)\right]L_3&=&0.\label{WLNPh}
\ea

Now, due to the extra symmetry of the non-trivial Killing tensor in these spacetimes, for the potentials  $L_N=L_N(v,u,\zeta,\ov\zeta)$ ($N=0-3$), we can always perform a separation of variables \cite{frolov} between those of the $(\boldsymbol{l}, \boldsymbol{n})$ subspace and those of the $(\boldsymbol{m}, \boldsymbol{\overline{m}})$ subspace in equation (\ref{Wave2}) .
Thus we can write 
\begin{equation}
L_N(v,u,\zeta,\ov\zeta)=L_{N_{1}}(v,u)L_{N_{2}}(\zeta,\ov\zeta). 
\end{equation}

\subsection{Initial Data}

On the 2D screen space of a null-hypersurface, the wave equation for $L_{N_{2}}(\zeta,\ov\zeta)$ reduces to 
\be\label{2dwave}
[(\delta+2\beta)\ov\delta+(\ov\delta+2\ov\beta)\delta-K]\{L_{N_{2}}(\zeta,\ov\zeta)\}=0
\ee
where $N_2$ runs from $0$ to $3$ and $K$ is a constant. 
This is the  2 D  part of a 3 D Laplace equation and the solutions of these are harmonic functions. In the case of a Kerr geometry, these are the spheroidal harmonics which reduce to spherical harmonics in the case of vanishing rotation. Here we recall an important theorem for the n-dimensional Laplace equation \cite{potential}:
\begin{thm}
The symmetries of the n-dimensional Laplace equation are exactly the conformal symmetries of  n-dimensional Euclidean space. 
\end{thm}
Therefore any solution of equation (\ref{2dwave}) (for example the spherical or spheroidal harmonics) are conformally symmetric and these can describe a 2 dimensional conformal Euclidean scalar field on the 2D screen space. \\

Now from equation (\ref{Wave2}), we can easily see that the equation for $L_{N_{1}}(v,u)$ on the $(\boldsymbol{l}, \boldsymbol{n})$ subspace reduces to 
\be\label{hyperbolic}
\left[-(\Delta+\mu+\ov\mu-\gamma-\ov\gamma)D-(D-\rho-\ov\rho)\Delta-K\right] L_{N_{1}}=0
\ee
which is a two dimensional hyperbolic partial differential equation, and the solution on the whole subspace can be uniquely found by specifying the initial data $L_{N_{1}}(v_0,u), DL_{N_{1}}(v_0,u)$ (or conversely  $L_{N_{1}}(v,u_0), \Delta L_{N_{1}}(v,u_0)$).\\

Therefore, the complete set of initial data ${\cal{I}}(\cal{S})$ on a null hypersurface ${\cal{S}}_1\equiv (u=u_0)$, required to solve equation (\ref{Wave2}) is given as
\ba\label{initial1}
{\cal{I}}({\cal{S}}_1)=\left\{L_{N_{1}}(u_0,v) L_{N_{2}}(\zeta,\ov\zeta), \Delta L_{N_{1}}(u_0,v)L_{N_{2}}(\zeta,\ov\zeta)\right\}
\ea

Obviously all the initial data in this set are not independent as they must obey the set of {\it Weyl-Lanczos} equations (\ref{WLNPa}-\ref{WLNPd}). Thus among the four potentials only two are needed to be specified on the null hypersurface ${\cal{S}}_1\equiv (u=u_0)$. However, once the initial data is specified then that uniquely determine the Weyl scalar in the bulk, $\Psi_2$, via equation (\ref{WLNPf}), and hence determine all the dynamical features of the gravity in bulk. Therefore there exist an one to one correspondence between any two the potentials at the null boundary and the non-zero Weyl scalar for type D spacetimes.

Similarly for the null hypersurface ${\cal{S}}_2\equiv (v=v_0)$, the required initial data will be
\ba\label{initial1}
{\cal{I}}({\cal{S}}_2)=\left\{L_{N_{1}}(u,v_0) L_{N_{2}}(\zeta,\ov\zeta), DL_{N_{1}}(u,v_0)L_{N_{2}}(\zeta,\ov\zeta)\right\}
\ea
 
Again in this case only two of the four potentials are independent as the initial data must obey (\ref{WLNPe}-\ref{WLNPh}). Subject to this initial data the Weyl scalar can be uniqely determined via equation (\ref{WLNPc}). In this case too, there exist an one to one correspondence between any two the potentials at the null boundary and the non-zero Weyl scalar.\\

    We can easily see that each element of the above sets of initial data, satisfies the two dimensional part of the Laplace equation (\ref{2dwave}) for each 2D screen space $(v_0,u_0)$ and hence represent a two dimensional conformal Euclidean field.  It is interesting that the conformal Euclidean scalar field lies on a two dimensional surface rather than a three dimensional one. This relates firstly to the fact that the intrinsic metric on any null surface is two dimensional and since we are specifying the initial data on a null surface, that must reside on the 2D screen space; and secondly there are symmetries on the 3-dimensional space (Killing tensors and vectors) that we have exploited in the above derivation.  In fact, from our knowledge of quantum thermodynamics of a Schwarzschild or Kerr black holes, this seems quite natural. Even in those cases the black hole entropy depends on the two dimensional surface area of a null event horizon rather than three dimensional volume.\\

 We can now state the  holographic principle for general relativity in this case:
	
\begin{quote}	\textbf{General Relativity  Holography}: \textit{On a null hypersurface $\cal{N}$ 
		in a vacuum Petrov type D spacetime with a non trivial Killing tensor (which describes all the astrophysical black hole spacetimes), the gravitational dynamics of four dimensional general relativity reduces to a classical euclidean conformal field theory in two dimensions for four scalar variables, with 
		data 
	$\{{\cal{I}}({\cal{S}}_1),{\cal{I}}({\cal{S}}_2)\}$ determing the 4-dimensional geometry.}
\end{quote}


That is, this is the conformal initial data on the null surface $\cal{N}$ that will determine the geometry  in the interior because of the existence and uniqueness theorems for the Einstein Field Equations in 4 dimensions. We can regard this as realizing Principles 1 and 2, as stated in Section 1,  for the case of classical General Relativity.

\section{Conclusion}
 We know in any vacuum spacetime the components of the Lanczos tensor obey massless scalar field equations. 
 If we consider the projection of this equation on the 2D screen space of a null hypersurface in a Kerr or Schwarzschild spacetime, then the Lanczos coefficients on this surface obeys a 2 dimensional Laplace equation, which is well known to be conformally invariant. A conformal field theory is a field theory which is invariant under conformal transformations\footnote{``Introducing conformal field theory'', notes by D Tong, DAMTP, Cambridge.}. Hence at least for black hole exteriors,  on a null hypersurface the gravitational dynamics of four dimensions reduce to a  Euclidean classical conformal field theory in two dimensions. This is a demonstration of the idea that even in classical general relativity, there exists an onto map from conformally invariant initial data at the null boundary to the dynamics of gravity in the bulk, Of course the inverse mapping does not exist because of the gauge condition (\ref{gauge}).
 This classical mapping has the great advantage that it will apply also for positive or negative cosmological constants, as well as for a vanishing cosmological constant. This is quite evident as the most general Petrov type D solutions are of  Plebanski and Demianski class (with a non-trivial Killing tensor) \cite{Kraetal80}, and the cosmological constant is just a parameter in this class of the metrics.\\
 
 We would like to emphasise here that our analysis is true for four dimensional spacetimes only. In higher dimensions, existence of the Lanczos tensor is not guaranteed and furthermore spacetimes admitting a non-trivial Killing tensor do not always imply separation of variables in the scalar wave equation. \\
 
 The extension to matter would presumably demand that the matter field added obeyed a conformal symmetry. Thus this result might extend to the case of a Maxwell field as a matter source term. Whether it holds for geometries with less symmetry is a matter for future investigation. Although the Lanczos equations are defined for general 4D spacetimes, one has to look carefully at the necessary conditions for the null initial data to have conformal symmetries.  \\
 
RG and GE thanks the South African National Research Foundation and GE thanks University of Cape Town research committee for financial  support.
 
 

\begin{thebibliography}{99}
 	\bibitem{Sus95} 	L Susskind ``The World as a Hologram'' \textit{J Math Phys} \textbf{36}: 6377-6396 [arXiv:hep-th/9409089].
 	
 	\bibitem{Bou02}
 	 R Bousso (2002) ``The Holographic Principle'' \textit{Reviews of Modern Physics} \textbf{74}: 825-874 [arXiv:hep-th/0203101].
	 
	\bibitem{Nat14}
	M Natsuume (2014)
	\textit{AdS/CFT Duality User Guide} Lecture Notes in Phyiscs 903 (Springer) [arXiv 1409.3575].
	
	
	\bibitem{Hub15} 	V E. Hubeny (2015)
	 ``The AdS/CFT Correspondence''  \textit{Class. Quantum Grav.} \textbf{32} 124010 	[arXiv:1501.00007 gr-qc]
	
 	 
 	 \bibitem{Ise13}  	 J Isenberg (2013)
 	 ``Initial Value Problem in General Relativity'' Chapter in \textit{The Springer Handbook of Spacetime}, ed. A. Ashtekar and V. Petkov. (Springer-Verlag)
 	 [arXiv:1304.1960 ].
 	 
 	 \bibitem{Sac62}
R K Sachs (1962) ``On the characteristic initial value problem in gravitational theory''  \textit{J.Math.Phys.} \textbf{3}: 908-914.
 	 
 	 \bibitem{Pen63}
 	 R Penrose (1963)
 	   ``Asymptotic properties of fields and space-times''  \textit{Phys.Rev.Lett}. \textbf{10}:66-68;  
 	   
 	   \bibitem{Pen80} R. Penrose (1980), 
``Null Hypersurface Initial Data for Classical Fields
of Arbitrary Spin and for General Relativitym'' 	   
 	    	   GRG Golden Oldie
 	    	   \textit{General Relativity and Gravitation} (1980) \textbf{12}:225-264.
 	   
	   \bibitem{Ray} R. A. d'Inverno1 and J. Stachel, ``Conformal two structure as the gravitational degrees of freedom in general relativity'' J. Math. Phys. {\bf 19}, 2447 (1978).
 	  
	   
	   \bibitem{Kerr} R P Kerr (1963) Phys. Rev. Lett. {\bf 11} 237.
 	   
 	   \bibitem{Ell676}
 	   G F R Ellis (1967)
 	   ``The dynamics of pressure-free matter in general relativity''
 	    	   \textit{Journ Math Phys} \textbf{8}:1171-1194.
 	   
 \bibitem{ElsEll96}
 	 H van Elst and G F R Ellis (1996)
 	    	   ``The covariant approach to LRS perfect fluid spacetime geometries''
 	   \textit{Classical and Quantum Gravity} \textbf{13}:1099.
 	   
 	   \bibitem{Lan38}
 C Lanczos (1938)
 ``A Remarkable Property of the Riemann-Christoffel Tensor in Four Dimensions''
 \textit{Annals of Mathematics}
 Second Series, \textbf{39}:842-850
 
 
 \bibitem{GolSac62} J N Goldberg and R K Sachs (1962) ``A theorem on Petrov types'' \textit{Acta Phys. Pol.} \textbf{22}:13?23.
 
 \bibitem{frolov} I D Novikov and V P Frolov, (1989) {\textit Physics of Black holes}  (Kluwer Academic, Boston).
 	   
 	   \bibitem{Rob89}
 	   M D Roberts
 	   (1989). ``Dimensional reduction and the Lanczos tensor''  \textit{Mod. Phys. Lett.} \textbf{A04}:2739. 
 \bibitem{OdBook} P O'Donnell  (2003) \textit{Introduction to 2-Spinors in General Relativity} (World Scientific, Singapore).
 
 \bibitem {Dolan}  P Dolan and C W Kim (1994) \textit{Proc. R. Soc.} \textbf{A 447}:577.	   
 
 \bibitem{Od} P O'Donnell (2004)   \textit{General Relativity and Gravitation} {\bf 36}:1415.

\bibitem {Bampi} F Bampi  and G Caviglia (1983) \textit{Gen. Rel. Grav.} \textbf{15}:375.
\bibitem {Ilge} R Illge (1988) \textit{Gen. Rel. Grav.} \textbf{20}: 551.
\bibitem{Edgar} S B Edgar and A H\"{o}glund (2000) \textit{Gen. Rel. Grav.} \textbf{32}:2307.
\bibitem{Maher}  W F Maher  and J D Zund (1968) \textit{Nuovo Cim.} \textbf{A 57}:638.
	   	   \bibitem{NewPen62}
 	    E Newman and R Penrose (1962)
 	    ``An Approach to Gravitational Radiation by a Method of Spin Coefficients''
 	    \textit{J. Math. Phys} \textbf{3}:566.
 	    
 	    \bibitem{Kraetal80}
 D Kramer, H Stephani, E Herlt, and M A H MacCallum (1980) \textit{Exact solutions of Einstein's field equations} (Cambridge: Cambridge University Press), Chapter 7.

\bibitem{EdHo97}
S B Edgar and  A. H\"{o}glund (1997) 	     
`The Lanczos potential for the Weyl curvature tensor: existence, wave equation and algorithms'' \textit{Proc Roy Soc} \textbf{A453}:835-851.
\bibitem{potential} A I Prilenko and E D Solomentsev (2001) "Potential theory". In  \textit{Encyclopedia of Mathematics} (Springer).

\bibitem{chandra} S Chandrasekhar, (1983) `{\it Mathematical theory of Black holes}' Clarendon Press, OUP, New York.

\bibitem{HawEll73} S W Hawking and G F R Ellis (1973) \textit{The Large Scale Structure of Space Time} (Cambridge: Cambridge University Press).

\bibitem{EllGos13}
G F R Ellis and R Goswami (2013)
``Variations on Birkhoff's theorem'' \textit{General Relativity and Gravitation} \textbf{45}: 2123-2142.

\bibitem{sommers} P Sommers (1976), Proc. Royal Soc. London A{\bf 349}, 309.
 
\end{thebibliography}
\end{document}